\documentstyle[preprint,aps,epsfig]{revtex}

\begin{document}
\tighten
\newcommand{\beq}{\begin{equation}}
\newcommand{\eeq}{\end{equation}}
\newcommand{\bea}{\begin{eqnarray}}
\newcommand{\eea}{\end{eqnarray}}
\newcommand{\ek}{\not{\varepsilon}}
\newcommand{\eek}{\vec{\varepsilon}}
\newcommand{\gp}{ {\cal G\/}_{F} }

\draft
\title{Properties of Dense Strange Hadronic Matter with Quark Degrees of
 Freedom}
\date{May 1, 2001}
\author{I. Zakout$^{1,2}$, H. R. Jaqaman$^{1}$, H. St\"ocker$^{2}$ and
W. Greiner$^{2}$}
\address{$^{1}$Department of Physics, Bethlehem University,
P.O. Box 9, Bethlehem, Palestinian Authority}
\address{$^{2}$Institut f\"ur Theoretische Physik,
J. W. Goethe Universit\"at, Frankfurt am Main, Germany}

\maketitle
\vspace{-10.0 cm}
\hfill Bethlehem University Preprint Phys-HEP001130
\vspace{ 10.0 cm}
\begin{abstract}
The properties of strange hadronic matter are studied in the context of the modified 
quark-meson coupling model using two substantially different sets of
hyperon-hyperon ($YY$) interactions.
The first set is based on the Nijmegen hard core potential model D with slightly attractive
$YY$ interactions. The second potential set is based on the recent $SU(3)$
extension 
of the Nijmegen soft-core potential NSC97 with strongly attractive $YY$ 
interactions which may allow for deeply bound hypernuclear matter. 
The results show that, for the first potential set, the $\Sigma$ hyperon does not appear at all in the bulk 
at any baryon density and for all strangeness fractions. 
The binding energy curves of the resulting $N\Lambda\Xi$ system vary smoothly with density 
and the system is stable (or metastable if we include the weak force). 
However, the situation is drastically changed when using the second set 
where the $\Sigma$ hyperons appear in the system at large baryon densities 
above a critical strangeness fraction.
We find strange hadronic matter undergoes a first order phase transition 
from a $N\Lambda\Xi$ system to a $N\Sigma\Xi$
for strangeness fractions $f_S>1.2$ and baryonic densities exceeding twice ordinary nuclear matter density. 
Furthermore, it is found that the system built of $N\Sigma\Xi$
is deeply bound. 
This phase transition affects significantly
the equation of state which becomes much softer and a substantial drop in energy density and pressure 
are detected as the phase transition takes place.
\end{abstract}

\narrowtext
\section{introduction}

Extremely dense strange quark matter with a strangeness fraction of 
order one and charge neutrality has been suggested to
be the absolutely stable form of matter at high densities (or at least 
to be metastable because of the weak interaction)
\cite{Bodmer71,Witten84,Farhi84,Berger87}.
Finite chunks of strange quark matter with large strangeness fractions, 
the so-called strangelets, are predicted to be more stable than 
normal nuclei
\cite{Farhi84,Gilson93,Greiner91,Barz90}.
Recent theoretical and experimental searches for strange quark matter 
can be found in Refs.\cite{SchA639,NagleA661}

On the other hand, metastable strange systems with strangeness fractions of 
order one and charge neutrality might also exist in
the hadronic phase at moderate values of density, between two 
and three times nuclear matter density.
The properties of metastable exotic multihypernuclear objects 
with $\Lambda$ and $\Xi$ hyperons 
reveal quite similar features as the strangelets proposed
as a unique signature for quark-gluon plasma formation in heavy ion
collisions \cite{SchC46}. The equilibrium between the quark and hadronic 
phases has been studied in \cite{Lee93}. 
Strange hadronic matter in bulk 
has been discussed by Glendenning\cite{Glend81}.

It is found that in an extended mean field theory, a large class of
bound multistrange objects formed from combinations of
$(N,\Lambda,\Xi)$ baryons are stable against strong decay.
The presence of filled $\Lambda$ orbitals blocks
the strong decay 
$\Xi N\rightarrow \Lambda\Lambda$\cite{SchAnn235}.
The maximal binding energy per baryon 
of $E_B/A\approx -21 \mbox{MeV}$ occurs
at a strangeness fraction or strangeness per 
baryon $f_s\approx$ 1.0-1.2, charge per baryon 
$f_q\approx$ -0.1-0.0 and baryon density 2.5-3 times that of ordinary
nuclei\cite{Schlett71}. 
It is comparable to that of hypothetically stable
strange quark matter (``strangelets''), which has a binding energy per
particle $E_B/A\approx$ -10 to -20 MeV.

The predicted phenomenon of metastability of strange hadronic matter 
and the actual values of the binding energy
depend specifically on the partly unknown hyperon potentials 
assumed in dense matter.
Some studies used basically Brueckner-Hartree-Fock (BHF) calculations 
with different Nijmegen soft core potentials
\cite{Ikeda85,Schulze98}.
Other studies\cite{Schlett71,Balberg94,WangA653} 
extend relativistic mean field theory (RMF)
from ordinary nuclei ($f_S=0$) to multistrange nuclei with ($f_S\neq 0$) 
with the attractive $YY$ interaction of the Nijmegen potential Model D
\cite{NaglesD15}.
                             
Recently, Stoks and Lee \cite{StoksC60} have studied strange 
hadronic matter using BHF theory and 
$G$ matrices for coupled baryon channels using 
an $SU(3)$ extension\cite{StoksC59,RijkenC59,VidanaC61} 
of the Nijmegen soft core NSC97 potential from the $S=$0,-1 sector 
into the unexplored $S=$-2,-3,-4 sector.
They have shown that $N\Lambda\Xi$ systems are only loosely
bound and that charge neutral strangeness-rich hadronic systems are 
unlikely to exist in nature.
Unfortunately their procedure is not self-consistent due to the 
constraint of equal hyperon fractions that they impose and hence 
does not give the minimum energy \cite{SchC62}.

Schaffner-Bielich and Gal \cite{SchC62} have carried out unconstrained RMF calculations 
and have found larger binding energies.
For small strangeness fractions $f_S\leq 1$, strange hadronic 
matter is mainly composed of
$N\Lambda\Xi$ and the calculated
binding energy closely follows that calculated by using the 
hyperon potentials 
of the earlier versions of the soft core Nijmegen potential.
For larger strangeness fractions $f_S\geq 1$, the calculated 
binding energy increases substantially
due to a phase transition into $N\Sigma\Xi$ dominated matter. 

In this work we use the quark-meson coupling Model (QMC) 
to study strange hadronic matter. The QMC model uses the 
quark degrees of freedom explicitly by coupling the
scalar $\sigma$ and vector $\omega$ mesonic fields directly to the up and 
down quarks inside the hadrons which are 
treated as non-overlapping MIT bags. It was first 
proposed by Guichon\cite{Guichon} and has been used in various nuclear 
calculations\cite{Blunden,Saito94}. More recently it has been  
modified to include a medium-dependent bag parameter coupled to
the scalar $\sigma$ field\cite{Jin,Jin2}.
This coupling is motivated by invoking the nontopological soliton model
for the nucleon\cite{soliton}. 
Quark deconfinement in hot nuclear matter 
and the phase transition from the hadronic phase to the quark-gluon plasma
have been studied in this model\cite{ZakC59,JaqamanG26}.
This modified quark-meson coupling model (MQMC)\cite{Jin,Jin2} has been extended
to the study of strange hadronic matter by introducing additional scalar 
$\sigma^*(s\overline{s})$
and vector $\phi(s\overline{s})$ mesonic fields which couple only to the
strange quark. 
This extended model has been used to study the properties of hot hypernuclear matter\cite{ZakC61}
and neutron star matter\cite{PalC60}.

We assume strange hadronic matter to consist of the baryon 
members of the SU(3) octet and decuplet. 
The octet is comprised of the nucleons 
and the $\Lambda$, $\Sigma$ and $\Xi$ hyperons while the 
decuplet contains the  $\Delta,\Sigma,\Xi$ and $\Omega$ baryons. The decuplet
baryons are however not found to have a significant contribution in the 
current calculations at zero temperature
and are henceforth dropped from our formalism.
We studied this system with two sets of hyperon-hyperon  ($YY$) potentials, 
the first set is determined from the Nijmegen hard-core 
potential Model D and the second set corresponds to the
potentials obtained in a recent SU(3) extension of the
Nijmegen soft-core potential Model NSC97. 
The differences between these sets are essentially attributable to the extremely 
attractive $\Sigma\Sigma$ and $\Xi\Xi$ interactions in the second set which allow the
possibility of deeply bound nuclear matter with hyperons.
We also consider these two potential models in the context 
of the QMC model with medium independent bag constant.

The current paper is organized
as follows. In section II we describe the details of the MQMC model,
in section III we show how we fit the various parameters in the
model and in section IV we present and discuss our results. Finally we
summarize our conclusions in section V.
 
\section{The MQMC Model}

In the QMC model, the quark field $\psi_{q}(\vec{r},t)$ inside a bag of radius $R_i$  
representing  a baryon of species $i$ satisfies the Dirac equation 
\begin{eqnarray}
\left[ i\gamma^{\mu}\partial_{\mu}- m_{q}^{0}
+(g_{q\sigma}\sigma-g_{q\omega}\omega_{\mu}\gamma^{\mu})
+(g_{q\sigma^{*}}\sigma^{*}-g_{q\phi}\phi_{\mu}\gamma^{\mu})
\right]\psi_{q}(\vec{r},t)=0,
\label{Dirac}           
\end{eqnarray}
where the quark is assumed to couple directly to the scalar and vector meson fields 
and $m_{q}^{0}$ is the current mass of a quark of flavor $q$.
The current quark masses are taken, for the up and down flavor quarks, 
to be $m_u=m_d=0$ while for the strange flavor $m_s=150\mbox{MeV}$ 
Inclusion of small current quark masses for the nonstrange flavors or
other values for the strange flavor leads only to 
small numerical refinements of the present results.  
In the mean field approximation 
the meson fields are treated classically and 
the space like components of the vector fields vanish for
infinite systems due to rotational invariance.
As a result $\omega_{\mu}\gamma^{\mu}=<\omega_0>\gamma^{0}=\omega\gamma^{0}$
and $\phi_{\mu}\gamma^{\mu}=<\phi_0>\gamma^{0}=\phi\gamma^{0}$.
The nonstrange (up and down) flavor quarks are coupled to the scalar 
$\sigma(550)$ and vector $\omega(780)$ mesons while the
strange flavor quarks are coupled to 
$\sigma^{*}(975)$ and $\phi(1020)$.

For a given value of  the bag radius $R_i$ for baryon species $i$
and the scalar fields
$\sigma$ and $\sigma^*$, the quark momentum
$x^i_q$ is  determined by the boundary condition of confinement which,
for quarks of flavor $q$ in a spherical bag, reduces to
$j_0({x^i_{q }})=\beta_{q} j_1({x^i_{q}})$, where
\begin{eqnarray}
\beta_{q}=
\sqrt{ \frac{
{\Omega^{i}_{q}}(\sigma,\sigma^*)
-R_i m^{*}_{q} }
{{\Omega^{i}_{q}}(\sigma,\sigma^*)
+R_i m^{*}_{q} }}.
\label{beta}            
\end{eqnarray}
We have defined the effective quark mass inside the bag as 
\begin{eqnarray}
m^{*}_{q}=m^{0}_{q}-g_{q\sigma}\sigma-g_{q\sigma^{*}}\sigma^{*},
\label{effmass}         
\end{eqnarray}
and
the effective quark energy is given by
\begin{eqnarray}
\Omega^i_q(\sigma,\sigma^*)/R_i=\sqrt{ (x_q/R_i)^{2}+{m_q^*}^{2} }.
\label{Omegnk}          
\end{eqnarray}
The bag energy  for baryon species $i$ is given by
\begin{eqnarray}
E^i_{bag}=
\sum^3_q
 n_{q}\frac{\Omega^i_{q}(\sigma,\sigma^*)}{R_i} 
- \frac{Z_i}{R_i}
+\frac{4\pi}{3}R_i^{3}  B_i(\sigma,\sigma^*),
\label{Ebag}         
\end{eqnarray}
where  $\frac{Z_i}{R_i}$ term is the zero-point energy of the quarks
and $B_i(\sigma,\sigma^*)$ is the medium-dependent bag parameter.
In the simple QMC model,
the bag parameter is taken as $B_{0}$ corresponding to its value 
for a free baryon. The medium effects are taken into account in the MQMC 
model \cite{Jin,Jin2}
by coupling the bag parameter to the scalar meson fields. 
In the present work we use the following generalized ansatz for 
the coupling of the bag parameter to the scalar fields
\begin{eqnarray}
B_i(\sigma,\sigma^*)=B_{0}\exp
\left[-4
\left(g^{\mbox{bag}}_{i\sigma}\sigma
+g^{\mbox{bag}}_{i\sigma^*}\sigma^*)/M_{i}\right)
\right]
\label{bagCon}          
\end{eqnarray}
with $g^{\mbox{bag}}_{i\sigma}$ and $g^{\mbox{bag}}_{i\sigma^*}$ as
additional parameters. Here it may be worth mentioning that in Ref. \cite{WangA653} 
the bag constant is coupled to the nonstrange $\sigma$ scalar field but not to $\sigma^{*}$.  

The spurious center-of-mass energy is subtracted to obtain 
the effective baryon mass\cite{Fleck} 
\begin{eqnarray}
M^{*}_{i}=\sqrt{{E^i_{bag}}^2 - <{p^{2}_{cm}}>^i},
\label{MNSTAR}          
\end{eqnarray}
where 
\begin{eqnarray}
<{p^{2}_{cm}}>^i=
\sum^3_q n_q x_q^2 / R_i^2.
\label{PCM}             
\end{eqnarray}
The bag radius $R_{i}$ for baryon species $i$ is obtained through the 
minimization of the baryon mass with respect to the bag 
radius\cite{Saito94}
\begin{eqnarray}
\frac{\partial M^{*}_{i}}{\partial R_{i}}=0.
\label{MNR}             
\end{eqnarray}
The zero-point energy parameters  $Z_i$ of Eq.(4) are used to fit the 
actual masses of the free baryons 
$M_i=939, 1116, 1189$ and $1315 \mbox{MeV}$
and are found to be 
$Z_i$=2.03, 1.814, 1.629 and 1.505 for the $N, \Lambda, \Sigma$ 
and $\Xi$ hyperons respectively, corresponding to a 
free baryon bag parameter $B_0=(188.1)^4 \mbox{MeV}^{4}$
and a free nucleon bag radius $R_0=0.6 \mbox{fm}$. 

The total energy density of cold infinite strange hadronic matter 
at finite baryon density $\rho_{B}$ is given by 
\begin{eqnarray} 
\varepsilon&=&
\sum_i \frac{\gamma_i}{(2\pi)^{3}} 
\int d^{3}k\sqrt{k^{2}+{M_{i}^{*}}^{2}}\theta(k^{i}_{F}-k)
+\frac{1}{2}m^{2}_{\omega} \omega^2
+\frac{1}{2}m^{2}_{\phi} \phi^2
+\frac{1}{2}m^{2}_{\sigma}\sigma^{2}
+\frac{1}{2}m^{2}_{\sigma^*}{\sigma^*}^{2} 
\label{Edensity}        
\end{eqnarray}
where the summation $i$ runs over the 8 members of the baryon octet which
reduces for symmetric hypernuclear matter to 4 species with the 
spin-isospin degeneracy factors
$\gamma_i=4,2,6$ and $4$ for $N$, $\Lambda$, $\Sigma$ and $\Xi$,
respectively and where 
$\theta$  is the step function  at 
the Fermi momentum $k^i_F$. 

The effective Fermi energy of baryon species $i$ is
given by  
\begin{eqnarray}
\epsilon^{*}_i(k^i_F)=
\sqrt{ {k^i_F}^{2}+{M^{*}_{i}}^{2}}+g_{i\omega}\omega
+g_{i\phi}\phi
\end{eqnarray}
which is equal to the chemical potential $\mu_i$
\begin{eqnarray}
\mu_i=B_i\mu_B+S_i\mu_S,
\end{eqnarray}
where $B_i$ and $S_i$ are the baryon and strangeness 
quantum numbers, respectively. 
The two independent chemical 
potentials $\mu_B$ and $\mu_S$ are obtained from the conservation 
of the total baryon density
\begin{eqnarray}
\rho_{B}=\frac{1}{(2\pi)^{3}}
\sum_i B_i \gamma_i \int d^{3}k \theta(k^{i}_{F}-k),
\label{rhoB1}         
\end{eqnarray}
and the total strangeness density
\begin{eqnarray}
\rho_{S}=\frac{1}{(2\pi)^{3}}
\sum_i S_i \gamma_i
\int d^{3}k \theta(k^{i}_{F}-k).
\label{rhoS1}         
\end{eqnarray}
The vector mean fields are determined  by
\begin{eqnarray}
\omega=\sum_i \frac{g_{i\omega}}{m^{2}_{\omega}} \rho_{i},
\label{vecomeg}        
\end{eqnarray}  
and
\begin{eqnarray}
\phi=\sum_i \frac{g_{i\phi}}{ m^{2}_{\phi}} \rho_{i},
\label{vecphi}         
\end{eqnarray}
where $g_{i\omega}$ and $g_{i\phi}$ are the meson-baryon coupling 
constants defined in Eqs. (18) and (19) below.

The pressure is the negative of the grand thermodynamic potential density
and is given by
\begin{eqnarray}
P&=&
\frac{1}{3}\sum_i\frac{\gamma_i}{(2\pi)^{3}}\int d^{3} k
\frac{k^{2}}{\epsilon_i^{*}}\theta(k^i_{F}-k)
+\frac{1}{2}m^{2}_{\omega}\omega^{2}
+\frac{1}{2}m^{2}_{\phi}\phi^{2}
-\frac{1}{2}m^{2}_{\sigma}\sigma^{2}
-\frac{1}{2}m^{2}_{\sigma^*}{\sigma^*}^{2}.
\label{pressurD}   
\end{eqnarray}  

\section{Fitting parameters for $YY$ potentials}


We assume that the  $\sigma$ and $\omega$ mesons 
couple only to the up and down quarks
while $\sigma^*$ and $\phi$ couple to the strange quark.
We thus set $g_{r\phi}=g_{r\sigma^{*}}=g_{s\sigma}= g_{s\omega}=0$ 
where $r$ refers to the up
and down flavors while $s$ denotes the strange flavor.
By assuming the $SU(6)$ symmetry of the simple quark model
we have the relations $g_{s\sigma^*}=\sqrt{2}g_{r\sigma}$ and 
$g_{s\phi}=\sqrt{2}g_{r\omega}$.
The coupling of each baryon species with the vector mesons 
is calculated by counting the constituent quarks
\begin{eqnarray}
g_{i\omega}=\sum_q^{3} g_{q\omega}=\sum_r g_{r\omega},
\end{eqnarray}
and 
\begin{eqnarray}
g_{i\phi}=\sum_q^3 g_{q\phi}=\sum_s g_{s\phi}.
\end{eqnarray}

With these assumptions the only free parameters left at our disposal are
the quark-meson coupling constants $g_{r\sigma}$ and $g_{r\omega}$
and the bag coupling constants
$g^{\mbox{bag}}_{i\sigma}, g^{\mbox{bag}}_{i\sigma^{*}}$ for the 4 
baryon species and these parameters 
are adjusted to fit  nuclear properties
as well as the extrapolated properties of hypernuclear matter.
The coupling constants of the scalar and vector mesons
to the nonstrange quarks
are taken as $g_{r\sigma}=1$ and $g_{r\omega}=2.705$ 
which together with a bag coupling constant
$g^{\mbox{bag}}_{N\sigma^{*}}=6.81$  yield a
binding energy of 16 MeV and a compressibility $K^{-1}_V$ of
289 MeV at the normal saturation
density $\rho_0=0.17 \mbox{fm}^{-3}$ of nuclear matter\cite{Jin2,ZakC59} .

Table I summarizes the values used in the current work for
the basic quark-meson coupling constants as
well as the two sets of
the coupling constants $g^{\mbox{bag}}_{i\sigma}$ and
$g^{\mbox{bag}}_{i\sigma^{*}}$
in the bag parameter $B_i(\sigma,\sigma^{*})$. These sets are chosen
to fit nuclear and hypernuclear properties. 
The parameters $g^{\mbox{bag}}_{i\sigma}$ are taken
to fit the hyperon potentials in nuclear matter:
\begin{eqnarray}
U^{(N)}_\Lambda (\rho_0)=-30 MeV, \nonumber \\
U^{(N)}_\Sigma (\rho_0)=+30 MeV, \nonumber \\
U^{(N)}_\Xi (\rho_0)=-18 MeV, \nonumber \\
\end{eqnarray}
where the hyperon potentials are defined by
\begin{eqnarray}
U^{(i)}_i=(M^{*}_i-M_i)+(g_{i\omega}\omega+g_{i\phi}\phi).
\end{eqnarray}
However, we make two different choices for the
constants $g^{\mbox{bag}}_{i\sigma^{*}}$.
In the first set, referred to hereafter as MQMC-I,
the medium constants $g^{\mbox{bag}}_{i\sigma^{*}}$ are 
adjusted so that the potential of a single hyperon 
embedded in a bath of $\Xi$ matter becomes
\begin{eqnarray}
U^{\Xi}_{\Xi}(\rho_0)=U^{\Xi}_{\Lambda}(\rho_0)=-40 MeV  
\end{eqnarray}
in accordance with the attractive hyperon-hyperon interaction
of the Nijmegen potential Model D\cite{SchAnn235,WangA653}.
Furthermore, we adopt the approximation
$U^{\Xi}_{\Xi}(\rho_0)\approx U^{\Xi}_{\Sigma}(\rho_0)$ 
to fit the medium constants.
The resulting $U^{(\Lambda)}_{\Lambda}(\rho_0/2)$
is about -20 MeV.

\begin{figure}
\centerline{\mbox{\epsfig{%
file=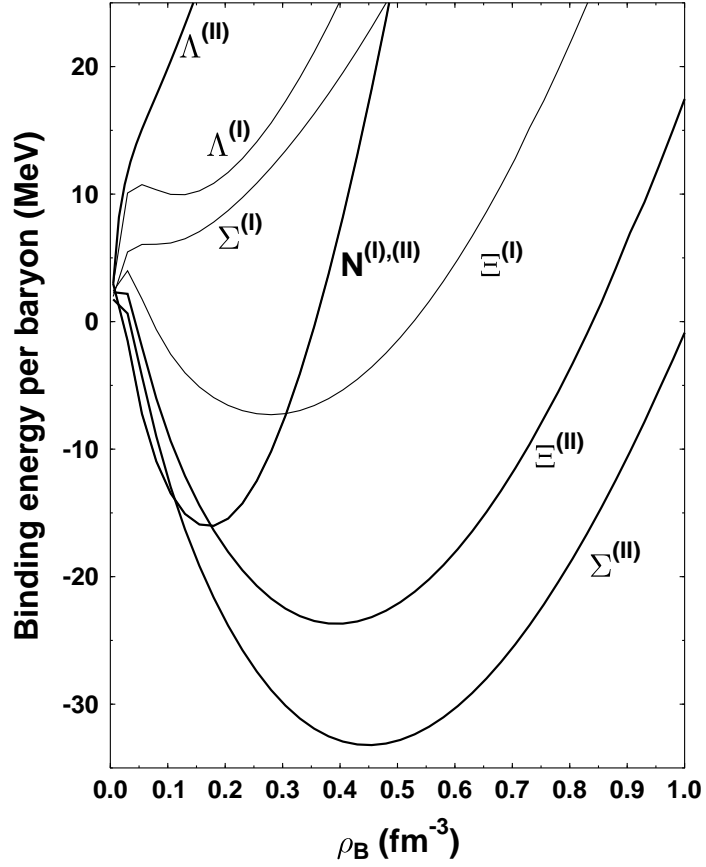,angle=0,width=0.8\linewidth}}}
\vspace{1truein}
\caption{Binding energy per nucleon $(N)$ in nuclear matter, compared
to the binding energy per hyperon $(\Lambda,\Sigma,\Xi)$ in its own
hyperonic matter. The hyperonic parameters where chosen to fit 
MQMC-I and II as discussed in the text. }
\label{fig_pot}
\end{figure}

In MQMC-II, we adopt the $YY$ interactions which occur
in the most recent $SU(3)$ extension of the Nijmegen soft-core
potential Model NSC97\cite{StoksC60,StoksC59,RijkenC59}.
The phenomenology in this model departs substantially from
that in MQMC-I.
In particular, the $\Sigma\Sigma$ and $\Xi\Xi$ interactions
are predicted to be highly
attractive in some channels, leading to bound states.
We have adjusted the bag constants 
$g^{\mbox{bag}}_{i\sigma^*}$ where $i=\Lambda,\Sigma,\Xi$ to
reproduce qualitatively the same  binding energy
curves of each hyperon species in its own hyperonic matter $B^i_i$
as those produced by the Model NSC97f\cite{StoksC60,SchC62}.
The resulting binding energy curves are shown in Fig.\ref{fig_pot}.
No binding occurs for $\Lambda$ hyperons  which are
already unbound by 8-10 MeV
at the rather low density $\rho_B=0.05 \mbox{fm}^{-3}$.
On the other hand,  $\Sigma$ matter is deeply bound at -33 MeV per baryon 
at ${\rho_\Sigma}_0$ which is twice as deep as ordinary nuclear matter, 
and $\Xi$ matter has an energy of -23 MeV per baryon at ${\rho_\Xi}_0$.

Furthermore, we have also considered the original (unmodified) QMC model where 
the bag constant is considered to be medium-independent and  
simply takes its free space value $B_0$. In analogy with the MQMC calculations we consider 
two sets of parameters. The QMC-I set reproduces the $YY$ potentials 
corresponding to the Nijmegen hard core potential model D as in MQMC-I. 
The scalar coupling  constants are taken as $g_{q\sigma}=5.31$ 
and $g_{s\sigma^*}=\sqrt{2}g_{q\sigma}$ 
while the vector ones are taken as
$g_{q\omega}$=1.471, 1.868, 5.148 and 1.662 and 
$g_{s\phi}$=0.0, 2.289, 0.0 and 3.017 for $N,\Lambda,\Sigma$ 
and $\Xi$, respectively.
The QMC-II set reproduces the $YY$  potentials corresponding to 
the Nijmegen soft-core potential NSC97 as in MQMC-II.
The scalar coupling constants are taken to be the same as in QMC-I while
the vector ones are taken as
$g_{q\omega}$=1.471, 2.942, 1.103 and 1.721 and 
$g_{s\phi}$=0.0, 4.160, 1.560 and 2.434 for $N,\Lambda,\Sigma$  
and $\Xi$, respectively.

\section{Results and Discussions}
The resulting energy per particle curves of each
baryon species $j$ in its own matter
$E^{(j)}_j$ are depicted in Fig. \ref{fig_pot} as a function of density
for models  MQMC-I and II.
For nucleons, the curves are the same in both models
since they fit the same nuclear matter properties.
In MQMC-I, the interactions  for $\Lambda$ and $\Sigma$ are
repulsive with $E^{\Lambda}_{\Lambda}$=+10 MeV 
and $E^{\Sigma}_{\Sigma}$=+5 MeV at $\rho_B=0.15 \mbox{fm}^{-3}$
for $\Lambda$ and $\Sigma$ hyperons respectively.
The $\Lambda$ hyperon has a shallow local minimum
at $\rho_B=0.13 \mbox{fm}^{-3}$ while
$E^{\Xi}_{\Xi}$ has an absolute minimum of $-7.5\mbox{MeV}$ per baryon 
at $\rho\cong 0.29\mbox{fm}^{-3}$.
On the other hand, for MQMC-II, the energy 
$E^{\Lambda}_{\Lambda}$ reaches $+20$ MeV 
already at a density $\rho_B=0.1\mbox{fm}^{-3}$.
This more repulsive potential can be attributed to the very weak underlying
$\Lambda\Lambda$ interaction in the extended NSC97 Model.
In contrast, the $\Sigma$ hyperon is deeply bound with a minimum of $-33$ MeV per baryon
at $\rho_B=0.46\mbox{fm}^{-3}$. 
Also $\Xi$ hyperon matter has a strong attractive potential
with a minimum of -23 MeV per baryon
at $\rho=0.39\mbox{fm}^{-3}$.
This shows that MQMC-II predicts that the $\Sigma\Sigma$   
and $\Xi\Xi$ interactions are highly attractive in some channels,
leading to bound states. 
It is also clear that a mixture of $\Sigma$ and $\Xi$ matter must be 
very deeply bound unless 
there is an overwhelmingly repulsive interaction
between the $\Sigma$ and $\Xi$ hyperons.

\begin{figure}
\centerline{\mbox{\epsfig{file=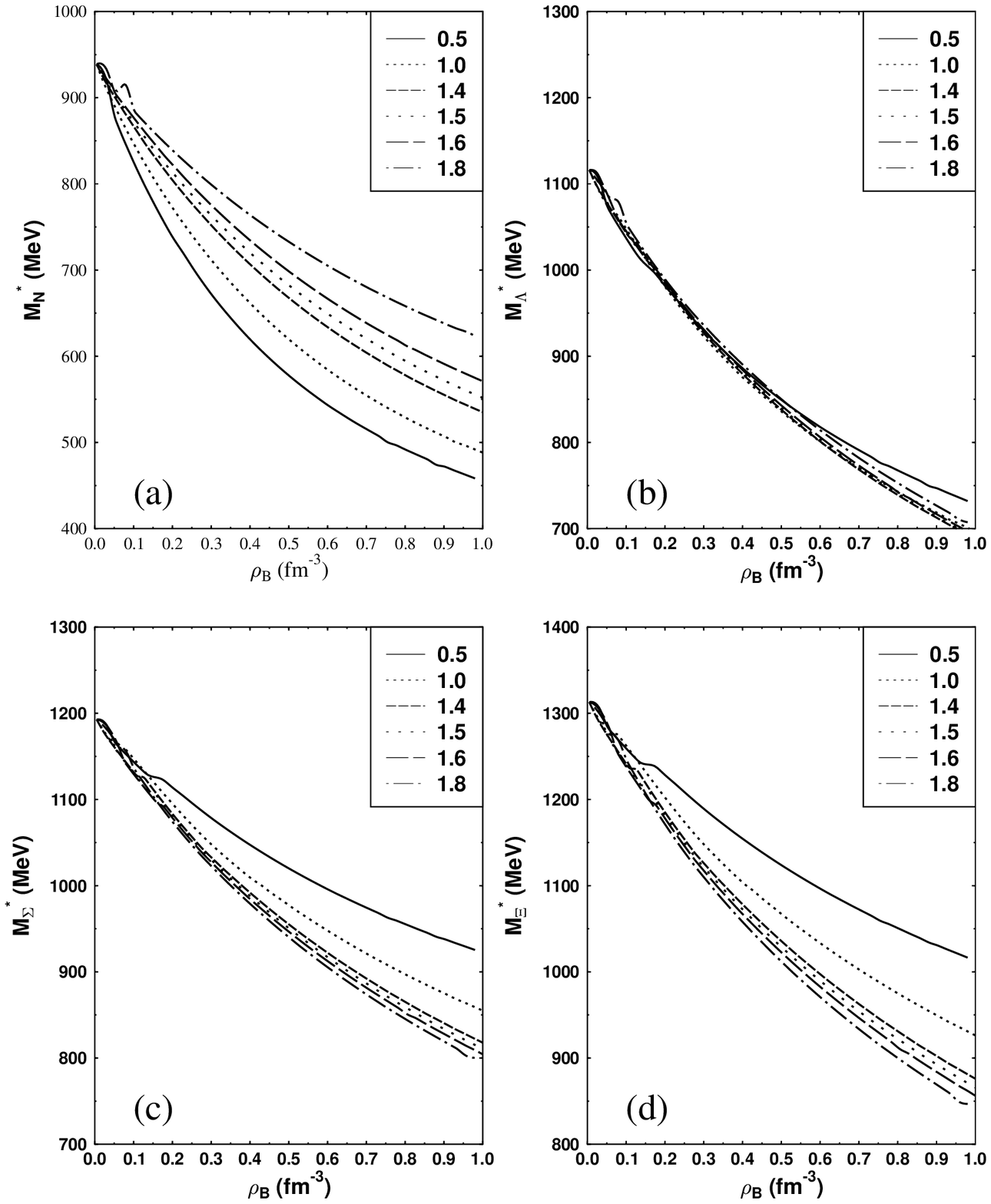,angle=0,width=0.8\linewidth}}}
\vspace{1truein}
\caption{The effective mass vs baryon density in MQMC-I for the $N,\Lambda,\Sigma,\Xi$ baryons.}
\label{fig_mas_a}
\end{figure}

\begin{figure}
\centerline{\mbox{\epsfig{file=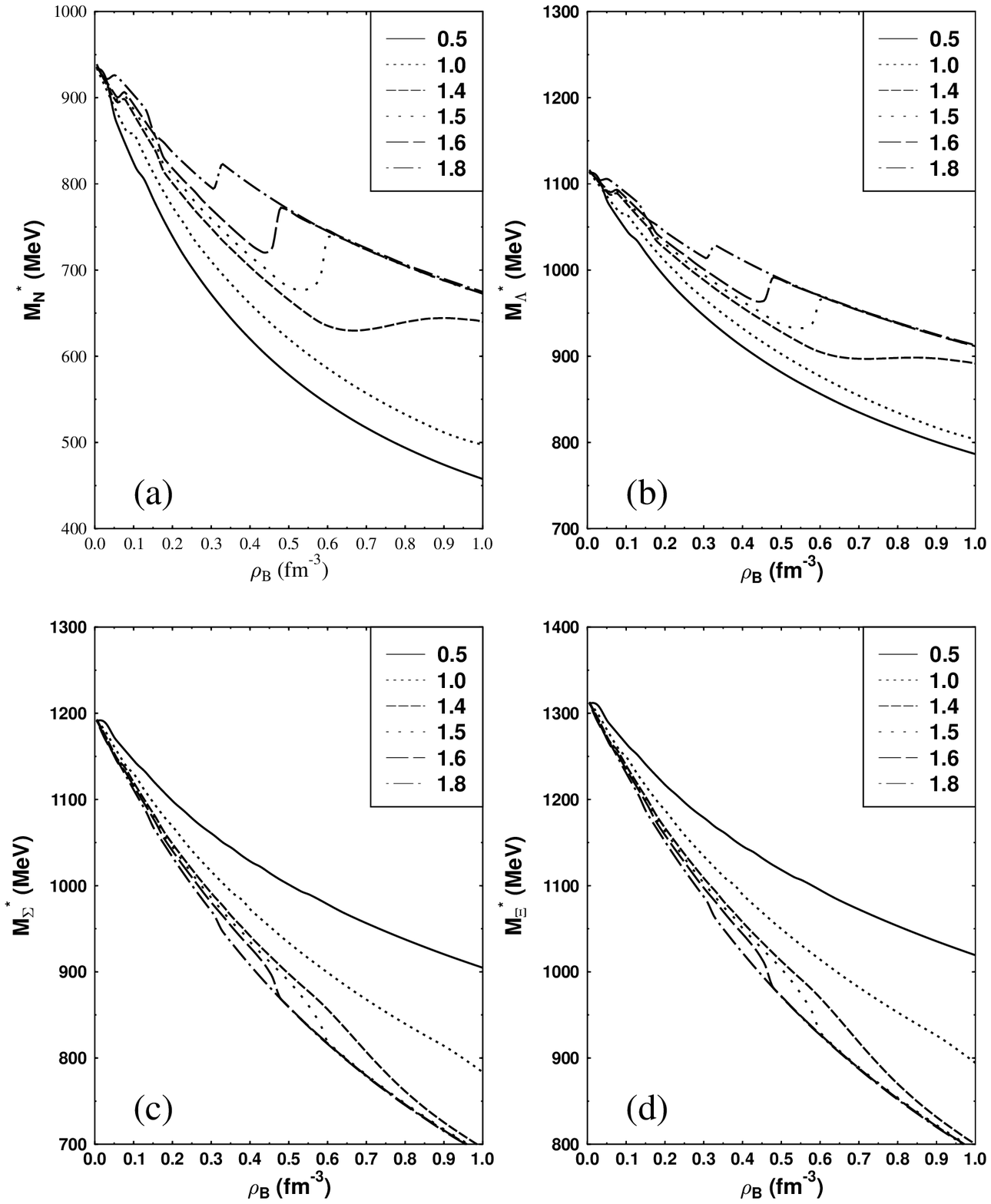,angle=0,width=0.8\linewidth}}}
\vspace{1truein}
\caption{The effective mass vs baryon density in MQMC-II for the $N,\Lambda,\Sigma,\Xi$ baryons.}
\label{fig_mas_b}
\end{figure}

Figs. \ref{fig_mas_a} and \ref{fig_mas_b} display the effective masses 
of the baryons $(N,\Lambda,\Sigma,\Xi)$ vs $\rho_B$ 
for several strangeness fractions
$f_S$ using MQMC-I and II, respectively.
In MQMC-I, the effective mass for each baryon species decreases
monotonically with density.
It is seen that as $f_S$ increases, $M^{*}_N$ increases 
while $M^{*}_{\Sigma}$ and  $M^{*}_{\Xi}$ decrease.
The $\Lambda$ hyperon effective mass however is only weakly sensitive 
to variations in $f_S$.
In MQMC-II, the situation is rather different, as can be seen 
from Fig. \ref{fig_mas_b}.
At first the effective masses of the nucleon and $\Lambda$ hyperon
decrease smoothly with $\rho_B$ for low strangeness fractions.
As $f_S$ exceeds 1.4, these effective masses jump suddenly to higher values
as $\rho_B$ reaches a critical value and then start to decrease
again monotonically for higher values of $\rho_B$.
The jump or discontinuity in the effective mass of the nucleon can be 
as high as $75$ MeV
for $f_S=1.5$ at a critical density of $0.57\mbox{fm}^{-3}$. 
The $\Lambda$ hyperon has a jump of around
40 MeV for $f_S=1.5$ at a critical density $\rho_B=0.57\mbox{fm}^{-3}$.
The effective masses of the $\Sigma$ and $\Xi$ hyperons
have the opposite behavior with
$M^{*}_{\Sigma}$ and $M^{*}_{\Xi}$ decreasing when 
the net strangeness fraction $f_S$ of the system increases.
Furthermore they suddenly jump to lower values 
as $\rho_B$ exceeds a critical density
for $f_S>1.4$.
It is interesting to note that in MQMC-II the various effective baryon masses 
become independent of $f_S$ for $f_S\ge 1.5$ 
and $\rho_B>0.60\mbox{fm}^{-3}$.
This behavior indicates that a phase transition takes place
in MQMC-II at the critical densities where the discontinuities in the effective
mass occur. The nature of this transition from the $N\Lambda\Xi$ phase to 
the $N\Sigma\Xi$ phase will become apparent in the following figures.

Figs. \ref{fig_rad_a} and \ref{fig_rad_b}
display the density dependence of the bag radius $R_i$ for the 
$N,\Lambda,\Sigma$ and $\Xi$ baryons for several strangeness fractions
$f_S$ using models MQMC-I and II, respectively.
The thick solid line indicates the limiting bag radius value
$R_{\mbox{exc}}=(3/4\pi\rho_B)^{1/3}$ where the excluded volume occupies
all space and
the assumption of nonoverlapping bags definitely breaks down.  
It is interesting to note here that the bag radius in MQMC
tends to increase with the increasing baryon density in contrary to the
ordinary QMC models where the bag radius tends to decrease slowly or 
saturate at a constant value.
This is because in MQMC the bag parameter
decreases with density which allows the bag to expand. This problem can be
cured by introducing quark-quark correlations \cite{JaqamanG26} which 
however are not included in the 
present calculations. Such correlations 
tend to make the bags shrink at higher densities so that the overlap 
assumption is not violated and $R_{\mbox{exc}}$ will have values 
much larger than those given by the thick line in Fig. \ref{fig_rad_a}. 
Finally it is observed that in MQMC-I the radius of the nucleon decreases 
while $R_{\Sigma}$ and $R_{\Xi}$ increase with increasing strangeness 
fractions while $R_{\Lambda}$ is almost unaffected by variations in $f_S$.

In contrast to MQMC-I where the bag radius increases monotonically with baryon 
density $\rho_B$ we notice from Fig. \ref{fig_rad_b} that in MQMC-II this variation 
has sudden jumps or discontinuities for $f_S\ge 1.4$.
For example, for a strangeness fraction $f_S=1.5$,
$R_N$ drops from $0.775\mbox{fm}$ to $0.725\mbox{fm}$ when the baryon
density reaches $\rho_B=0.55\mbox{fm}^{-3}$.
The shrinking in the nucleon bag size for $f_S> 1.4$
takes place before the breakdown in the nonoverlap assumption even in the current
calculations where the quark-quark correlations are not included. 
Furthermore, the phase transition tends to occur at lower and lower 
densities as $f_S$ increases
and the bag radius becomes independent of $f_S$.  
The same behavior is found for the radius of the $\Lambda$ hyperons but
the situation is a quite different for the $\Sigma$ and $\Xi$ hyperons.
$R_\Sigma$ ($R_\Xi$) both increase with respect to $f_S$ and both
jump to a higher value at the critical baryon density
when the phase transition takes place for $f_S$ exceeding 1.4. With 
$f_S=1.5$, $R_{\mbox{exc}}$ is reached at $\rho_B=0.50\mbox{fm}^{-3}$
for $\Sigma$ and $\Xi$ hyperons indicating that, even with the 
neglect of quark-quark correlations, MQMC-II is valid up to 
$\rho_B=0.50\mbox{fm}^{-3}$ for such large finite strangeness fractions.

\begin{figure}
\centerline{\mbox{\epsfig{file=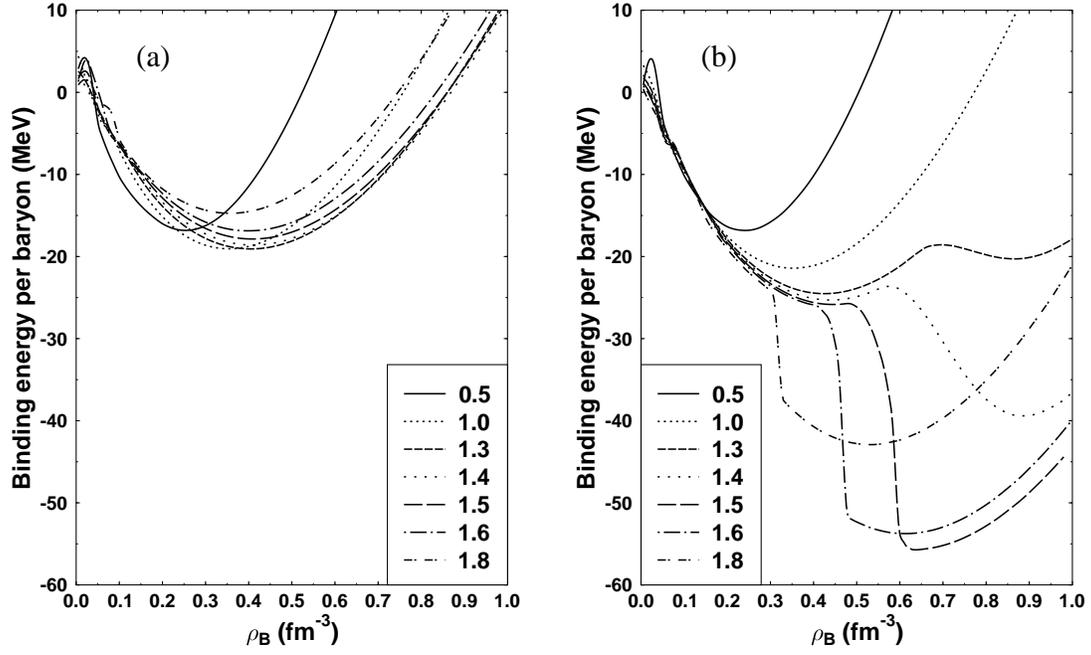,angle=-90,width=1.0\linewidth}}}
\vspace{1truein}
\caption{The binding energy per baryon of strange hadronic matter  for
various strangeness fractions in (a) MQMC-I and (b) MQMC-II.}
\label{fig_bind}
\end{figure}

Fig. \ref{fig_bind} displays the binding energy per baryon 
for strange hadronic matter 
vs the baryon density $\rho_B$  for various
strangeness fractions $f_S$. 
It is noted in Fig.\ref{fig_bind}(a) that the energy curves
for MQMC-I have only one global minimum for every value of the
strangeness fraction $f_S$.
At first, the depth of the minimum increases as 
$f_S$ increases until it reaches a strangeness fraction 
of about 1.3 after which the depth of the minimum decreases
again for higher values of $f_S$ and approaches
that of purely $\Xi$ matter for $f_S=2$.
It is found that in MQMC-I strange hadronic matter 
is built out always from a mixture of $N,\Lambda$ and $\Xi$ baryons.
Hence each energy curve has only one global minimum. 

Fig. \ref{fig_bind}(b) indicates that the situation for MQMC-II
is drastically different.
For low values of $f_S$ each curve still has a single minimum whose 
depth increases with the increase in
the strangeness fraction. However, 
when $f_S$ reaches about $1.3$, a second minimum 
appears at a rather high density $\rho_B=0.86\mbox{fm}^{-3}$.
Moreover, this second energy minimum quickly deepens  
and reaches $-39$ and $-56$ MeV for $f_S=1.4$  and 1.5, respectively.
The appearance of this second minimum can be attributed to
the increasingly dominant presence of 
the $\Sigma$ hyperons in the system which seems to undergo
a first order phase transition from $N\Lambda\Xi$
and $N\Xi$ to $N\Sigma\Xi$ strange hadronic matter but
at a rather higher baryonic density.
For values of the strangeness fraction $f_S$ greater than 1.5   
the potential well depth starts to decrease again as the $\Xi$
starts to dominate and the energy curves gradually approach that
for pure $\Xi$ matter for  $f_S=2$.

\begin{figure}
\centerline{\mbox{\epsfig{file=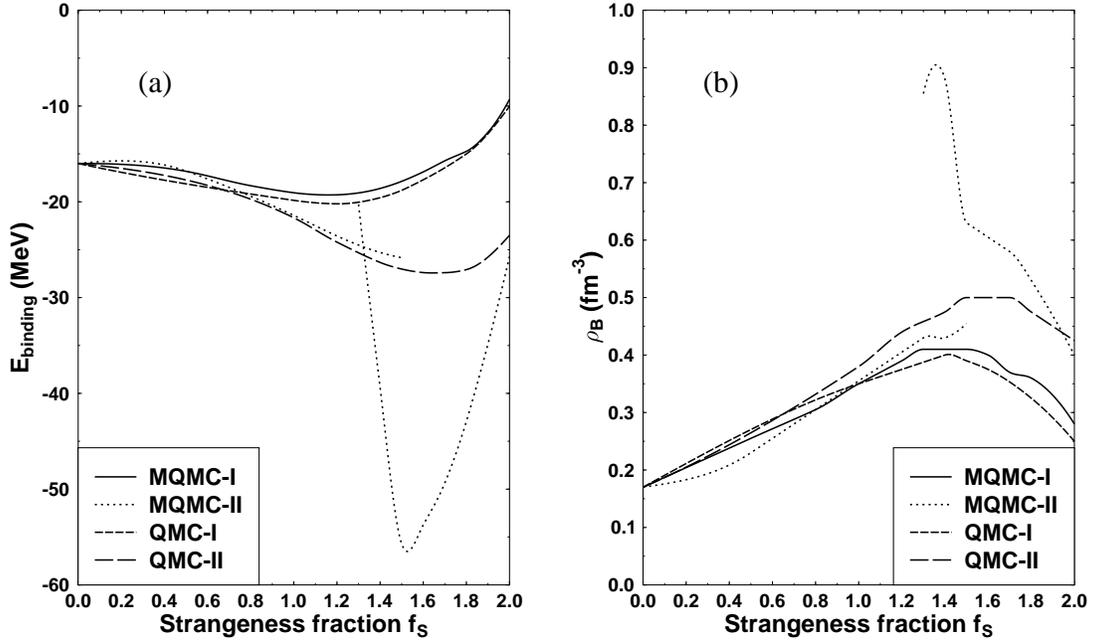,angle=-90,width=1.0\linewidth}}}
\vspace{1truein}
\caption{Comparison of (a) the depth and (b) the location of the
energy minimum of strange hadronic
matter as functions of the strangeness fraction $f_s$ in MQMC-I
and MQMC-II as well as in QMC-I and QMC-II.}
\label{fig_min}
\end{figure}

Fig. \ref{fig_min} displays the depth
of the energy minimum and the corresponding baryonic density
as functions of the strangeness fraction $f_S$.
It is evident that the depth of the minimum in MQMC-I increases 
with the strangeness fraction
until the curve has a maximum binding energy of $19\mbox{MeV}$
at $f_S=1.2$. 
The binding energy magnitude 
decreases again as $f_S$ increases to form pure
$\Xi$ matter at $f_S=2.0$
with a binding energy of 10 MeV. 
As can be seen from Fig.\ref{fig_min}(b) the locations of these
minima appear at higher densities as $f_S$
increases until it reaches $f_S=1.2$ where it hovers around
$\rho_B=0.40\mbox{fm}^{-3}$
for $f_S=1.2-1.6$. 
Beyond $f_S=1.6$, the location of the energy minimum 
moves to lower densities.

The situation for MQMC-II is rather similar to that of 
MQMC-I for $f_S<1.3$,
but it is essentially different for $f_S\geq1.3$. 
When the strangeness fraction reaches $f_S=1.3$,
a second minimum in the binding energy appears
at $\rho_B=0.86\mbox{fm}^{-3}$.
Furthermore, the first minimum disappears at $f_S=1.5$ 
and a transition to the second minimum (the $N\Sigma\Xi$ phase)
occurs due to the reaction $N\Xi\rightarrow\Sigma\Sigma$.
The upper and lower branches of the MQMC-II curve in 
Fig.\ref{fig_min}(b) thus refer to two different phases with the
discontinuity in the density of the system indicating
it is a first order transition.
The binding energy reaches a maximum of 56 MeV
for $f_S=1.5$ and $\rho_B=0.64\mbox{fm}^{-3}$ and then 
decreases to 25.7 MeV as $f_S$ increases to $f_S=2$
where the second minimum disappears since the system
is now composed solely of $\Xi$ hyperons.
We also display in Fig. \ref{fig_min} the results for the QMC-I and QMC-II models 
which are seen to be very similar to those found in MQMC-I  with 
a single stable $N\Lambda\Xi$ phase. In particular, unlike MQMC-II, 
QMC-II does not lead to a stable $N\Sigma\Xi$ system.
Finally it is worth mentioning that, due to the shrinking of the bag radius
with increasing baryon density, the nonoverlap assumption survives up to 
$\rho_B>1.1\mbox{fm}^{-3}$. 

\begin{figure}
\centerline{\mbox{\epsfig{file=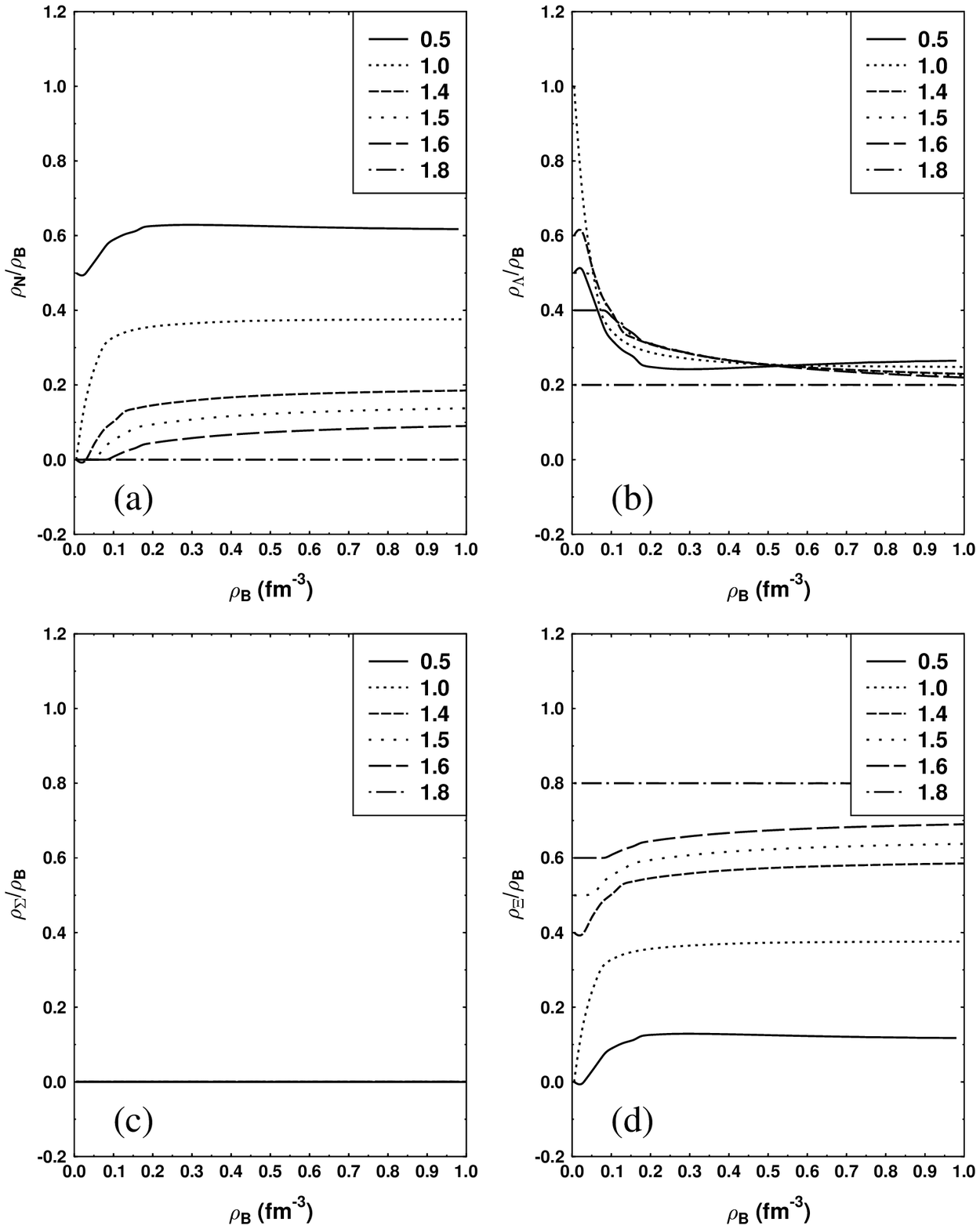,angle=0,width=0.8\linewidth}}}
\vspace{1truein}
\caption{Fractional density $\rho_i/\rho_B$ vs. $\rho_B$ for each
baryon species in MQMC-I with different strangeness fractions.
(a) $N$, (b) $\Lambda$, (c) $\Sigma$
and (d) $\Xi$.}
\label{fig_dens_a}
\end{figure}

\begin{figure}
\centerline{\mbox{\epsfig{file=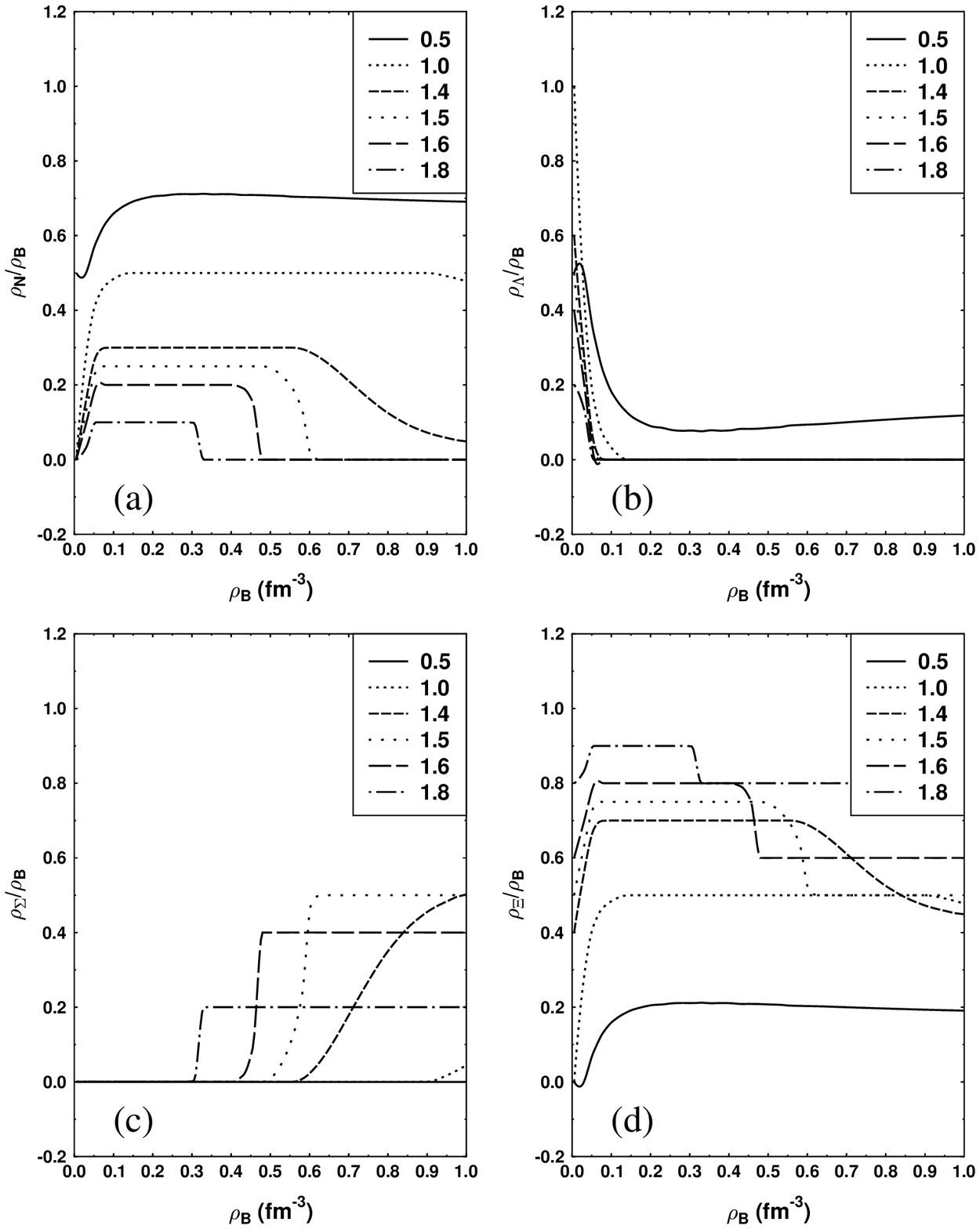,angle=0,width=0.8\linewidth}}}
\vspace{1truein}
\caption{Fractional density $\rho_i/\rho_B$ vs. $\rho_B$ for each
baryon species in MQMC-II with different strangeness fractions.
(a) $N$, (b) $\Lambda$, (c) $\Sigma$
and (d) $\Xi$.}
\label{fig_dens_b}
\end{figure}

Figs. \ref{fig_dens_a} and \ref{fig_dens_b} display the fractional density
of each baryon species $\rho_i/\rho_B$ in strange hadronic matter
versus the total baryon density $\rho_B$ for several values of $f_S$ 
for Models I and II, respectively.
It is seen that in MQMC-I the $\Sigma$ hyperon does not 
appear at all while the $\Lambda$ has a substantial contribution
for low densities $\rho_B\leq 0.1\mbox{fm}^{-3}$. 
The $\Lambda$ contribution, however, drops substantially for higher
densities to saturate at a constant value around 
$\rho_{\Lambda}/\rho_{B}=0.20-0.25$
for $\rho_B\geq 0.2\mbox{fm}^{-3}$ and $f_S=0.5-1.8$.
It also disappears for $f_S=0$ (nuclear matter) and $f_S=2.0$
(pure $\Xi$ matter).
The contribution of the nucleons $\rho_N/\rho_B$ decreases monotonically
from one to zero as $f_S$ increases from zero to two.
For $f_S\geq 1$ the nucleons comprise a surprisingly small fraction
of the total number of baryons for small densities  but their contribution
increases rapidly and saturates at higher densities.
The variation of the contribution  of the $\Xi$ hyperons
with density
is similar to that of the nucleons but with 
$\rho_{\Xi}/\rho_B$ increasing from zero to one as $f_S$ increases
from zero to two.

Fig. \ref{fig_dens_b} shows however that the situation is significantly 
different in MQMC-II, with the $\Sigma$ hyperons appearing above  
some critical density and their contribution rapidly saturating
at higher densities. The fractional density 
$\rho_{\Sigma}/\rho_B$ takes its
maximum value of about $50\%$ at $f_S=1.5$. 
The $\Lambda$ hyperon contribution is large at low baryon densities
$\rho_B<0.10$ but practically disappears for higher densities
for $f_S\geq1$.
The  fractional densities for the $N$'s and $\Xi$'s in MQMC-II are similar
to those in MQMC-I except for $f_S\geq1.2$ where there is a depletion
at high densities caused by the appearance of the $\Sigma$ hyperons
in the system. The nucleons even disappear
completely at high densities for $f_S\geq1.5$.

\begin{figure}
\centerline{\mbox{\epsfig{file=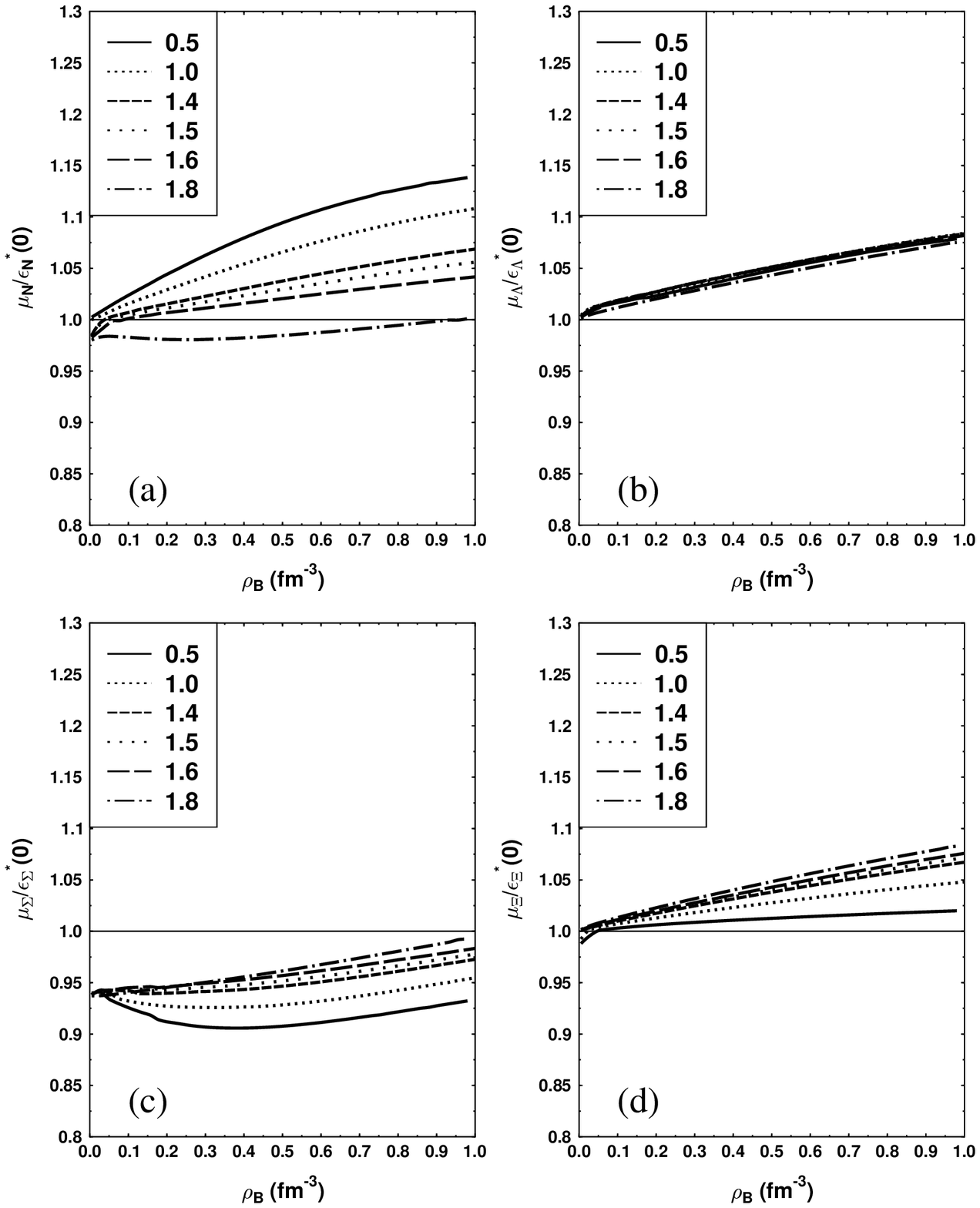,angle=0,width=0.8\linewidth}}}
\vspace{1truein}
\caption{The threshold production ratio $\mu_i/\epsilon^{*}_i(k_F=0)$
for each baryon species and for different strangeness fractions in MQMC-I.
(a) $N$, (b) $\Lambda$, (c) $\Sigma$ and (d) $\Xi$.}
\label{fig_thres_a}
\end{figure}

\begin{figure}
\centerline{\mbox{\epsfig{file=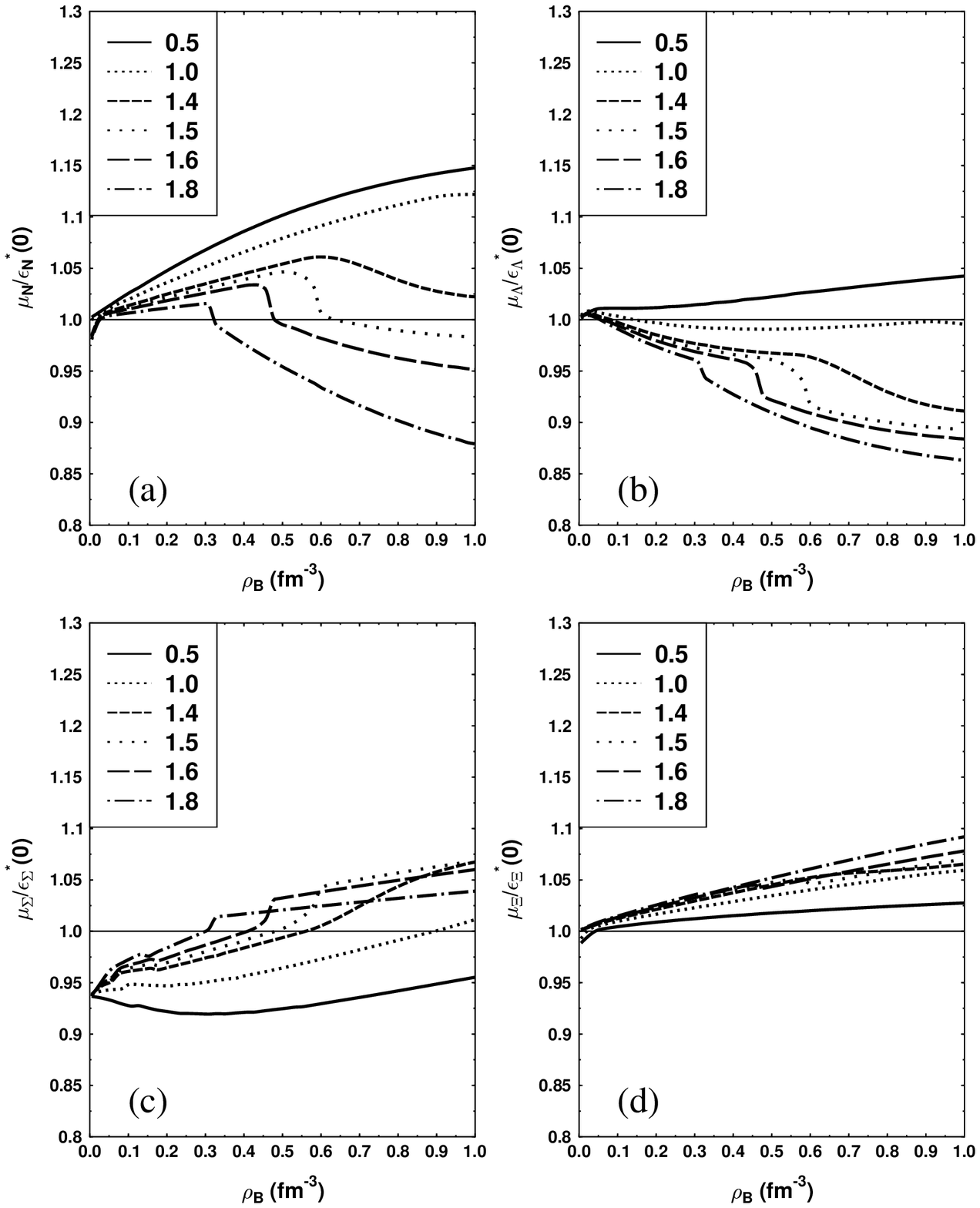,angle=0,width=0.8\linewidth}}}
\vspace{1truein}
\caption{The threshold production ratio $\mu_i/\epsilon^{*}_i(k_F=0)$
for each baryon species and for different strangeness fractions in MQMC-II.
(a) $N$, (b) $\Lambda$, (c) $\Sigma$ and (d) $\Xi$.}
\label{fig_thres_b}
\end{figure}

Figs. \ref{fig_thres_a} and \ref{fig_thres_b}
display $\mu_i/\epsilon^{*}_{i}(0)$ vs $\rho_B$
$\epsilon^{*}_{i}$ is the Fermi energy and $\mu_i$ 
is the chemical potential, eqs. (11) and (12). 
If the ratio $\mu_i/\epsilon^{*}_{i}(k_F=0)<0$ baryons of 
species $i$ cannot exist in the system. 
The baryon species first appears
with zero fermi momentum $k_F=0$
when this ratio reaches one and then takes
a finite fermi momentum as the ratio exceeds one.  
>From Fig. \ref{fig_thres_a} 
it is evident that in MQMC-I the $\Sigma$ hyperons
cannot be produced at all in strange hadronic matter.
Furthermore, $N$ production is increasingly suppressed 
while $\Xi$ production is enhanced as $f_S$ increases.

The situation is quite different in MQMC-II, Fig. \ref{fig_thres_b}, 
where the $\Sigma$ hyperons appear
in the system  for strangeness fractions $f_S\geq1.0$.
They first appear at $\rho_B\geq0.85$ for $f_S=1.0$ and
as  $f_S$ increases they appear at lower baryon densities. 
However, the situation is opposite for $\Lambda$
which is produced abundantly for low strangeness 
fractions $f_S<1.0$ and then disappears 
when $f_S$ exceeds $1.0$. 
The contribution of the nucleons
decreases significantly for $f_S>1.3$  and they disappear 
completely from the system at $f_S=1.5$ 
when the baryon density exceeds $\rho_B=0.65 \mbox{fm}^{-3}$.
Furthermore, the nucleons disappear at lower baryon densities as the
strangness fraction $f_S$ increases. It is thus seen that 
$N$ production in MQMC-II is influenced drastically by 
the transition from the $N\Lambda\Xi$ phase to the $N\Sigma\Xi$ phase.
On the other hand,  $\Xi$ hyperon production 
in MQMC-II is not much different from that in MQMC-I.

\begin{figure}
\centerline{\mbox{\epsfig{file=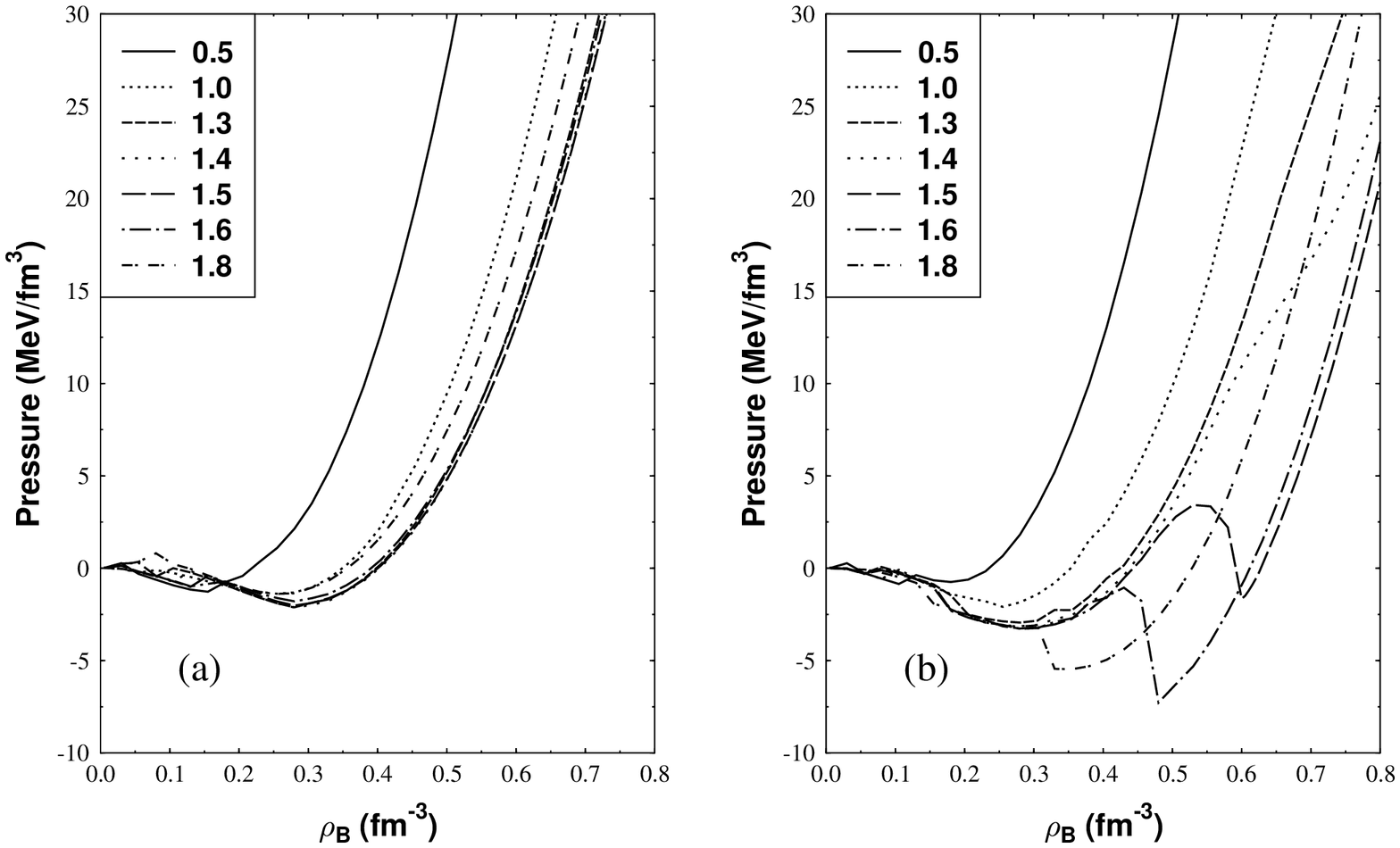,angle=-90,width=1.0\linewidth}}}
\vspace{1truein}
\caption{Pressure of strange hadronic matter vs baryon density 
for different strangeness fractions in (a) MQMC-I and (b) MQMC-II.}
\label{fig_press}
\end{figure}

Finally, Fig.\ref{fig_press} displays the pressure vs the
baryon density $\rho_B$ for Models I and II.
MQMC-I has the normal van der Waal-type loop at small densities
reflecting the  gas to liquid phase transition. 
In MQMC-II the pressure has an additional sudden and 
significant drop  at high densities  $\rho_B>0.55 \mbox{fm}^{-3}$
for strangeness fractions $f_S\geq 1.5$.
This dramatic drop in the pressure is due to
the transition from the $N\Lambda\Xi$ phase to the $N\Sigma\Xi$
phase in strange hadronic matter and probably provides an experimental clue 
for it.

\section{Summary and conclusion}

In the present work, we have studied, within the context of the MQMC model, 
the properties of strange hadronic matter
composed of the SU(3) octet baryons $N,\Lambda,\Sigma$ and $\Xi$
using two different models for the hyperon-hyperon interactions.  
It is interesting to note here that the medium dependent bag constant
for each baryon species reflect more realistic self interaction with
the scalar mean fields which are corresponding to the nonlinear terms
of the scalar mean fields in the normal RMF calculations.

MQMC-I is designed to mimic the consequences of the Nijmegen
hard-core potential model D \cite{NaglesD15,SchAnn235,Balberg94}.
It is nevertheless constrained by $\Lambda$
and $\Xi$ nuclear phenomenology, and by a few $\Lambda\Lambda$
hypernuclear species  reported to date.
It accounts  more realistically for the attractive $\Lambda\Lambda$ and $N\Xi$ interactions,
but ignores altogether $\Sigma$ hyperons.
On the other hand, MQMC-II is designed to generate qualitatively similar baryon
potentials to those obtained in the BHF approximation from the SU(3) extensions 
of the Nijmegen soft-core NSC97 potential models.
The NSC97 model has been tuned up to reproduce certain characteristics of $\Lambda$ hypernuclei,
particularly its version NSC97f.
It yields particularly attractive $\Xi\Xi$, $\Sigma\Sigma$ and $\Sigma\Xi$
interactions. 

In MQMC-I, the $\Sigma$ hyperons do not appear at all and the $\Lambda$ hyperons have 
a substantial contribution at low densities for medium strangeness fractions $f_S=0.5-1.8$. 
However this contribution drops quickly and saturates for $\rho_B/\rho_0 \geq 1$ 
at a constant fraction density slightly larger than $20\%$.
This model produces smooth binding energy curves over the entire range of 
$f_S$ and $\rho_B$ and forms (meta-)stable
strange hadronic matter in the bulk.

Due to the  exceptionally strong attractive interactions among the
$\Sigma$ and $\Xi$ hyperons in MQMC-II,
the $N\Lambda\Xi$- and $N\Xi$-dominated strange hadronic matter that exists for $f_S\leq 1.3$
is replaced by $N\Sigma\Xi$-dominated matter for $f_S\geq 1.3$, with binding
energies per baryon reaching as much as 56 MeV for $f_S=1.5$.
The equation of state becomes much softer and seems 
to allow deeply bound hyperonic matter. The system built out of $N\Sigma\Xi$ forms
a particularly (meta-)stable state of strange hadronic matter in the bulk.
When the first order phase transition takes place above some
critical baryon density and strangeness fraction, 
the $\Sigma$ hyperons appear and their contribution saturates rapidly to
a constant fraction density that takes its maximum value of about
$50\%$ at $f_S=1.5$. It is expected that such a phase transition may 
drastically change the global features of neutron stars\cite{Hanauske}.

Our results for MQMC-II are qualitatively similar to those of 
Schaffner-Bielich and Gal \cite{SchC62} where the same $YY$ 
potentials of MQMC-II (what they call Model N) are 
employed within RMF theory but with only hadronic degrees of freedom.
There are however some important quantitative differences. Whereas in 
Ref. \cite{SchC62} it is found that the second energy minimum, corresponding to $N\Sigma\Xi$
dominated matter, becomes more stable than the first minimum for $f_S\geq 1$, our current 
results indicate that this takes place at higher strangeness fractions $f_S\geq 1.3$.
Furthermore, the binding energy well for the $N\Sigma\Xi$ system has a maximum depth 
of 56 MeV in the present work
while Ref. \cite{SchC62} finds the largest depth to be 78 MeV. In both calculations the 
largest binding for the $N\Sigma\Xi$ system
takes place at  $f_S\approx 1.5$. 
The fact that the phase transition takes place at larger 
strangeness fractions may make it more difficult 
to observe experimentally, if it occurs at all.

Finally, quark-quark correlations and excluded
volume effects may  become important at high baryon densities 
in particular when the bags start to overlap 
at baryon densities $\rho_B=2-3 \rho_0$.
We shall consider these effects on 
the properties of strange hadronic matter in our future work.  

\acknowledgments
Financial support by  the Deutsche Forschungsgemeinschaft through 
grant GR 243/51-2 is gratefully acknowledged.
We also thank A. Gal for helpful discussions.


%
\begin{table}
\caption{Fitting parameters }
\label{states}
\begin{tabular}[b]{cccccccccc}
Fit set & $g_{q\sigma}$ & $g_{q\omega}$ & $g^{\mbox{bag}}_{N\sigma}$ &  
$g^{\mbox{bag}}_{\Lambda\sigma}$ & $g^{\mbox{bag}}_{\Lambda\sigma^{*}}$ & 
$g^{\mbox{bag}}_{\Sigma\sigma}$ & $g^{\mbox{bag}}_{\Sigma\sigma^{*}}$ & 
$g^{\mbox{bag}}_{\Xi\sigma}$ & $g^{\mbox{bag}}_{\Xi\sigma^{*}}$ \\
\hline
MQMC-I  &1.0& 2.705 & 6.81 & 4.22 & 5.45 & 1.63 & 7.26 & 2.27 & 9.12 \\
MQMC-II &1.0& 2.705 & 6.81 & 4.22 & 0.0 & 1.63 & 10.28 & 2.27 & 10.17 \\
\end{tabular}
\end{table}

%
\begin{figure}
\centerline{\mbox{\epsfig{file=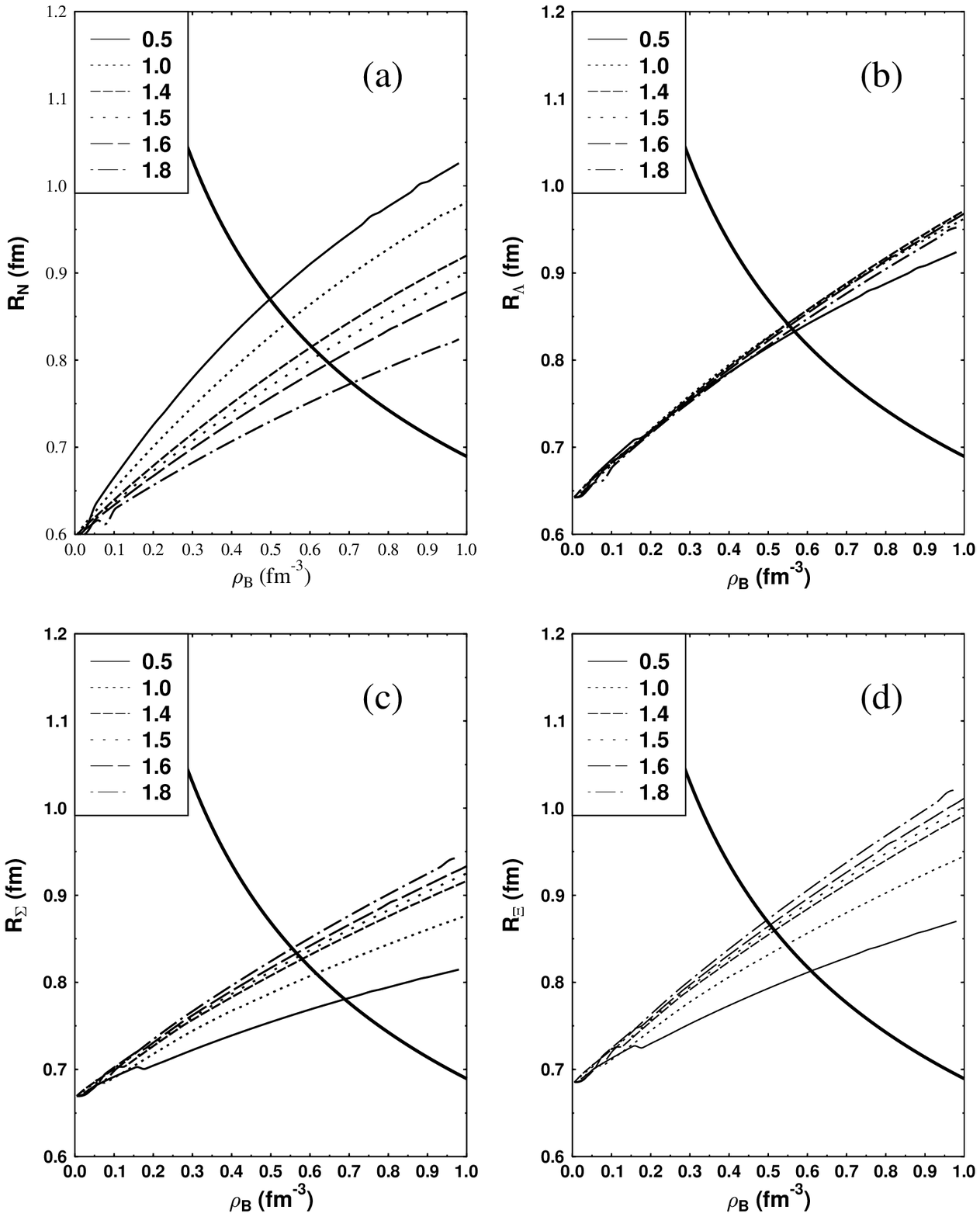,angle=0,width=0.8\linewidth}}}
\vspace{1truein}
\caption{The bag radius vs baryon density in MQMC-I for the $N,\Lambda,\Sigma,\Xi$ baryons.}
\label{fig_rad_a}
\end{figure}
\begin{figure}
\centerline{\mbox{\epsfig{file=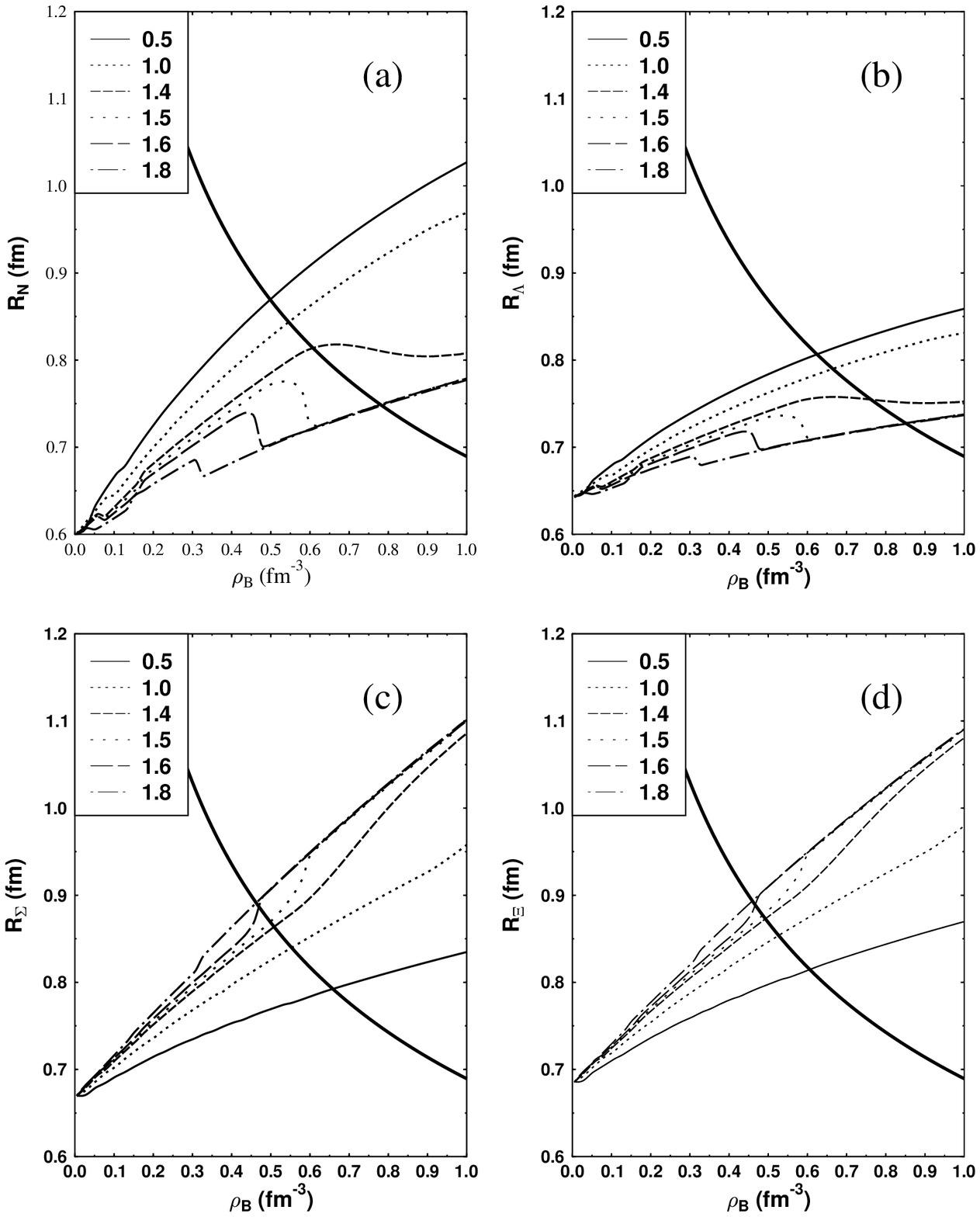,angle=0,width=0.8\linewidth}}}
\vspace{1truein}
\caption{The bag radius vs baryon density in MQMC-II for the $N,\Lambda,\Sigma,\Xi$ baryons.}
\label{fig_rad_b}
\end{figure}

\begin{references}
\bibitem{Bodmer71}
A. R. Bodmer, Phys. Rev D {\bf 4}, 1601 (1971).

\bibitem{Witten84}
E. Witten, Phys. Rev. D {\bf 30}, 272 (1984).

\bibitem{Farhi84}
E. Farhi and R. L. Jaffe, Phys. Rev. D {\bf 30}, 2379 (1984).

\bibitem{Berger87}
M. S. Berger and R. L. Jaffe, Phys. Rev. C {\bf 35},213 (1987)

\bibitem{Gilson93}
E. P. Gilson and R. J. Jaffe, Phys. Rev. Lett. {\bf 71}, 332 (1993).

\bibitem{Greiner91}
C. Greiner and H. St\"ocker, Phys. Rev. D {\bf 44}, 3517 (1991)

\bibitem{Barz90}
H. W. Barz, B. Friman, J. Knoll and H. Schulz, Phys. Lett. {\bf B242}, 328 (1990).

\bibitem{SchA639}
J. Schaffner-Bielich, Nucl. Phys. {\bf A639}, 443c (1998).

\bibitem{NagleA661}
J. L. Nagle for the E864 Collaboration, Nucl. Phys. {\bf A661}, 185c
(1999).

\bibitem{SchC46}   
J. Schaffner, C. Greiner and H. Sto\"ocker, Phys. Rev. C {\bf 46}, 322 (1992).


\bibitem{Lee93}
K. S. Lee and U. Heinz, Phys. Rev. D {\bf 47}, 2068 (1993).

\bibitem{Glend81}
N. Glendenning, Phys. Rev. C {\bf 23}, 2757 (1981).

\bibitem{SchAnn235}  
J. Schaffner, C. B. Dover, A. Gal, D. J. Millener, C. Greiner and H. St\"ocker,
Ann. Phys. (N.Y.) {\bf 235}, 35 (1994).

\bibitem{Schlett71} 
J. Schaffner, C. B. Dover, A. Gal, C. Greiner, and H. St\"ocker, Phys. Rev. Lett. {\bf 71}, 1328 (1993).

\bibitem{Ikeda85}
K. Ikeda, H. Bando, and T. Motoba, Prog. Theor. Phys. Suppl. {\bf 81}, 147 (1985).

\bibitem{Schulze98}
H.-J. Schulze, M. Baldo, U. Lambardo, J. Cugnon, and A. Lejeune, Phys. Rev. C 
{\bf 57} 704 (1998). 

\bibitem{Balberg94}
S. Balberg, A. Gal and J. Schaffner, Prog. Theor. Phys. Suppl. {\bf 117}, 325 (1994).

\bibitem{WangA653}
P. Wang, R. K. Su, H.Q. Song, and L.L. Zhang, Nucl. Phys. {\bf A653}, 166 (1999).

\bibitem{NaglesD15} 
M.M. Nagels, T.A. Rijken, and J.J. de Swart, Phys. Rev. D {\bf 15}, 2547 (1977).

\bibitem{StoksC60}
V. G. J. Stoks and T.S.H. Lee, Phys. Rev. C {\bf 60}, 024006 (1999)

\bibitem{StoksC59}
V. G. J. Stoks and Th. A. Rijken, Phys. Rev. C {\bf 59}, 3009 (1999).

\bibitem{RijkenC59} 
Th. A. Rijken, V. G. J. Stoks, and Y. Yamamoto, Phys. Rev. C {\bf 59}, 21 (1999).

\bibitem{VidanaC61}
I. Vida\~na, A Polls, A. Ramos, M. Hjorth-Jensen, and V.G.J. Stoks, Phys. Rev. C {\bf 61}, 025802 (2000).

\bibitem{SchC62} J. Schaffner-Bielich and A. Gal, 
Phys. Rev C {\bf 62} 034311 (2000).

\bibitem{Guichon}
P. A. M. Guichon, Phys. Lett. B {\bf 200}, 235 (1988).
\bibitem{Blunden}
P. G. Blunden and G. A. Miller, Phys. Rev. C {\bf 54}, 359 (1996).
\bibitem{Saito94}
K. Saito and A. W. Thomas, Phys. Lett. B {\bf 327}, 9 (1994).
\bibitem{Jin}
X. Jin and B. K. Jennings, Phys. Rev. C {\bf 54}, 1427 (1996);
Phys. Lett. B {\bf 374}, 13 (1996).
\bibitem{Jin2}
X. Jin and B. K. Jennings, Phys. Rev. C {\bf 55}, 1567 (1997);  
H. M\"uller and B. K. Jennings, Nucl. Phys.  {\bf A 626}, 966 (1997);
H. M\"uller, Phys. Rev. C {\bf 57}, 1974 (1998).
\bibitem{soliton}
M. K. Banerjee, Phys. Rev. C {\bf 45}, 1357 (1992);
V. K. Mishra, Phys. Rev. C {\bf 46}, 1143 (1992);
E. Naar and M. C.  Birse, J. Phys. G. {\bf 19}, 555 (1993)
\bibitem{ZakC59}
I. Zakout and H. Jaqaman,  Phys.Rev. C {\bf 59} 962 (1999);{\it ibid}
{\bf 59} 968 (1999).
\bibitem{JaqamanG26}
I. Zakout and H. R. Jaqaman, J.Phys. G {\bf 26} 1095 (2000).
\bibitem{ZakC61}
I. Zakout, H. R. Jaqaman, S. Pal, H. St\"ocker and W. Greiner,
Phys. Rev. C {\bf 61} 055208 (2000).
\bibitem{PalC60}
S. Pal, M. Hanauske, I. Zakout, H. St\"ocker and W. Greiner
Phys. Rev. C {\bf 60} 015802 (1999).
\bibitem{Fleck}
S. Fleck, W. Bentz, K. Shimizu, and K. Yazaki, Nucl. Phys. {\bf A510},
731 (1990).


\bibitem{Hanauske} 
J. Schaffner-Bielich, M. Hanauske, H. St\"ocker and W. Greiner, astro-ph/0005490.
\end{references}
\end{document}